\documentclass{article}
\usepackage{longtable}
\usepackage{booktabs,tabularx,siunitx}
\usepackage{PRIMEarxiv2}
\usepackage{tikz}
\usetikzlibrary{arrows.meta,positioning,shapes.geometric}
\usepackage[utf8]{inputenc} 
\usepackage[T1]{fontenc}    
\usepackage{hyperref}       
\usepackage{url}            
\usepackage{booktabs}       
\usepackage{amsmath}        
\usepackage{amsfonts}       
\usepackage{amssymb}        
\usepackage{nicefrac}       
\usepackage{microtype}      
\usepackage{lipsum}
\usepackage{fancyhdr}       
\usepackage{graphicx}       
\graphicspath{{media/}}     
\usepackage{enumitem}       

\title{A Costing Framework for Fusion Power Plants
\thanks{\textit{\underline{Citation}}: 
\textbf{Authors. Title. Pages.... DOI:000000/11111.}} 
}

\author{
  Simon Woodruff \\
  Woodruff Scientific Ltd \\
  1 Kings Meadow \\
  Oxford\\
  \texttt{\{Simon Woodruff\}simon@woodruffscientific.com} \\
}

\begin{document}
\maketitle

\begin{abstract}
This paper summarizes and consolidates fusion power-plant costing work performed in support of ARPA-E from 2017 through 2024, and documents the evolution of the associated analysis framework from early capital-cost-focused studies to a standards-aligned, auditable costing capability. Early efforts applied ARIES-style cost-scaling relations to generate Nth-of-a-kind (NOAK) estimates and were calibrated through a pilot study with Bechtel and Decysive Systems to benchmark balance-of-plant (BOP) costs and validate plant-level reasonableness from an engineering, procurement, and construction (EPC) perspective. 
Subsequent work, informed by Lucid Catalyst studies of nuclear cost drivers, expanded the methodology to treat indirect costs explicitly and to evaluate cost-reduction pathways for non-fusion-island systems through design-for-cost practices, modularization, centralized manufacturing, and learning. 
As ARPA-E’s fusion portfolio expanded, these methods were applied across BETHE and GAMOW concepts (and select ALPHA revisits), including enhanced treatment of tritium handling and plant integration supported by Princeton/PPPL expertise. 
In 2023 the capability was refactored to align with the IAEA--GEN-IV EMWG--EPRI code-of-accounts lineage, while key ARIES-derived scaling relations were replaced by bottom-up subsystem models for dominant fusion cost drivers (e.g., magnets, lasers, power supplies, and power-core components) coupled to physics-informed power balances and engineering-constrained radial builds. 
These developments were implemented in the spreadsheet-based Fusion Economics code (FECONs) and released as an open-source Python framework (pyFECONs), providing a transparent mapping from subsystem estimates to standardized accounts and a consistent computation of LCOE. 
\end{abstract}

\section{Introduction}
Economic credibility is now a central requirement for fusion energy: as concepts mature from physics demonstrations toward integrated pilot plants, developers, funders, and policymakers require cost estimates that are transparent, comparable across architectures, and traceable to underlying technical assumptions. The ARPA-E support work summarized here was motivated by precisely this need: to move beyond one-off, non-comparable fusion costing exercises toward a repeatable programmatic capability that can identify dominant cost drivers, test cost-reduction strategies, and enable like-for-like comparisons with advanced fission and other low-carbon generation options.

\subsection*{Foundation and guiding philosophy}
The work builds on three complementary foundations: (i) the ARIES family of fusion power plant studies, which couple physics-informed power balances, engineering-constrained radial builds, conceptual plant layouts, and high-level cost scalings; (ii) ARPA-E’s 2017 pilot study with Bechtel and Decysive Systems, which benchmarked and sanity-checked fusion plant capital costs from an EPC standpoint; and (iii) Lucid Catalyst analyses of nuclear cost drivers, which emphasize the importance of indirect costs, construction methodology, modularization, and learning in determining achievable \$/kW outcomes. 
Across these efforts, the guiding workflow remained consistent: begin with a physics-informed power balance, translate performance requirements into a feasible radial build subject to engineering constraints, size dominant driver systems (e.g., magnets, lasers, power supplies), and then translate thermal performance and layout into BOP equipment and buildings, with conventional plant subsystems anchored primarily to NETL baselines where appropriate. 

\subsection*{Program evolution: from pilot studies to portfolio support}
The program evolved through distinct stages. In 2017, the immediate objective was to establish an initial fusion costing workflow suitable for ARPA-E needs and to calibrate ARIES-derived outputs against EPC experience; this phase highlighted sensitivities to layout, buildings, and major equipment sizing and established a credible baseline for BOP-related costs. 
In 2019, the emphasis broadened to separate and scrutinize BOP and (especially) indirect costs, and to identify actionable pathways for reducing non-power-core costs via standardized layouts, modularization, centralized manufacturing, and learning effects---a key shift driven by the recognition that non-fusion-island costs often dominate total plant cost. 
During 2022--2023, the capability matured into a portfolio-scale support function for BETHE and GAMOW awardees (and selected ALPHA revisits), providing concept-consistent analyses across diverse private-sector architectures while improving realism in tritium handling and fuel-cycle assumptions through PPPL’s operational and safety expertise. 

\subsection*{Standards alignment and tool release}
By 2023, ARPA-E’s growing portfolio and the need for cross-concept comparability motivated a refactor toward a standardized, auditable, and extensible framework. This refactor aligned the costing structure with the IAEA (2001), GEN-IV EMWG (2007), and EPRI (2024) code-of-accounts lineage, and prioritized replacement of legacy ARIES scalings with bottom-up subsystem models for major fusion-unique drivers, coupled to explicit power-balance and radial-build calculations (including multi-fuel and blanket option coverage).
The resulting capability was implemented in FECONs (spreadsheet) and released as pyFECONs (open-source Python), emphasizing transparency, traceability, and extensibility as a reference implementation for credible fusion costing and consistent LCOE evaluation.

\subsection*{Scope and contributions}
Accordingly, this paper (i) documents the evolution of the ARPA-E fusion costing methodology from ARIES-derived scalings to bottom-up subsystem models, (ii) summarizes how portfolio-scale application sharpened identification of dominant cost drivers and improved tritium/fuel-cycle realism, and (iii) presents the standardized account structure and implementation philosophy embodied in the FECONs/pyFECONs tools, enabling auditable and like-for-like cost comparisons across fusion concepts and against other generation technologies.

\section{Background}
\label{sec:background}

This work adopts a deliberately conservative and well-documented costing lineage, drawing on established power-plant cost-accounting methodologies that have been developed and refined within the nuclear sector. Specifically, the cost-category structure and economic accounting conventions used here are informed by (i) International Atomic Energy Agency (IAEA) guidance on economic evaluation of nuclear power plants, (ii) the Generation IV International Forum (GIF) Economics Modeling Working Group (G4EMWG) guidelines, and (iii) the synthesis and interpretation of these methods in Rothwell's \emph{Economics of Nuclear Power} framework \cite{iaea-2001,gif-emwg-2007,Rothwell2015}. The purposeful use of these contemporary, externally vetted methodologies improves transparency and facilitates reproducibility for researchers who wish to perform comparable analyses.

A key departure from many prior fusion costing studies is the explicit reorganization of the capital cost accounts. Direct costs are separated into Category 10 (pre-construction) and Category 20 (construction), while indirect and capitalized ancillary costs are grouped into Categories 30--60 (formerly Categories 91--98 in some legacy nuclear accounting structures). This structure retains substantial continuity with established practice---for example, the Category 20 construction accounts remain closely aligned with conventional plant breakdowns---while adding or clarifying elements that are particularly important for modern fusion development programs (e.g., explicit treatment of a digital twin and contingency). Additionally, the treatment of decommissioning is modified to improve accounting clarity: rather than embedding decommissioning in the LCOE in a manner that can be difficult to audit, decommissioning costs are represented as a capitalized indirect cost, enabling clearer attribution, sensitivity studies, and discussion of the drivers of end-of-life obligations.

Although the accounting structure is derived from fission-centered references, the cost drivers for fusion plants can differ substantially in the heat island, fuel cycle, remote handling, and replacement schedule of major components. Those fusion-specific departures are therefore made explicit throughout the paper at the subsystem level. Where possible, we adopt the GEN-IV subcategory descriptions from the G4EMWG guidance, editing them to capture fusion-specific hardware and operational realities without losing the comparability benefits of a standardized taxonomy.

Our costing philosophy is to (i) provide recent cost-basis information for each cost category wherever possible, (ii) state the cost basis and its provenance directly in the text, and (iii) provide references that enable users of the accompanying models and code to trace assumptions back to primary sources. We are also judicious in terminology. In some contexts, the term ``reactor'' carries connotations that may hinder broader adoption of fusion energy systems; accordingly, we use the term ``heat island'' where it more precisely conveys function without unnecessary connotation. We recognize that terminology will evolve, and we aim for internal consistency and clarity in this edition.

\begin{figure}[h!]
\centering
\includegraphics[scale=0.45]{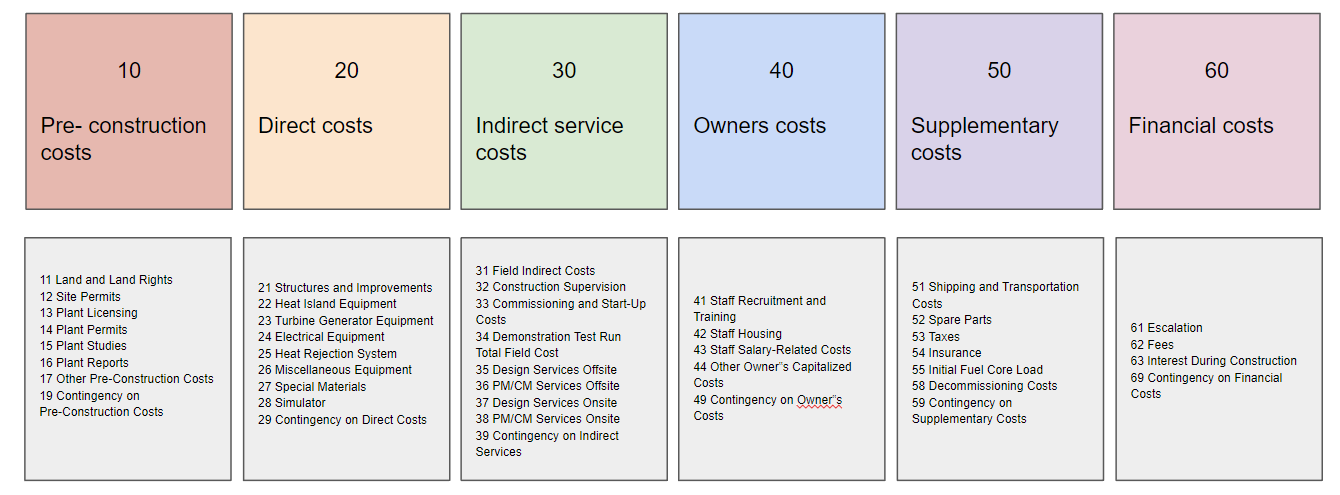}
\caption{Cost categories used in this work. The capital-cost taxonomy separates direct costs into Category 10 (pre-construction) and Category 20 (construction), and groups indirect and capitalized ancillary costs into Categories 30--60.}
\label{fig:costcategories}
\end{figure}

\subsection{Foundational documents influencing the costing methodology}
\label{subsec:foundational_docs}

The GEN-IV Economics Modeling Working Group (G4EMWG) guidelines provide a comprehensive structure for economic modeling of advanced nuclear energy systems \cite{gif-emwg-2007}. The guidance systematically defines cost categories spanning pre-construction activities through annualized financial costs, and it provides a granular breakdown of capitalized costs (direct costs, capitalized indirect service costs, owner’s costs, and supplementary costs), as well as recurring costs (operations and maintenance, fuel, and financing). In addition to defining the taxonomy, the guidelines emphasize disciplined cost accounting practices that support transparent budgeting, economic feasibility evaluation, and consistent comparisons across designs and deployment contexts.

Rothwell’s treatment of nuclear power economics provides an integrative foundation for interpreting these accounting conventions within the broader context of investment decision-making \cite{Rothwell2015}. It emphasizes the relationship between capital cost determination, uncertainty and risk, and the financial structure of large-scale energy infrastructure. The framework is particularly relevant for fusion, where high capital intensity and technology novelty increase the importance of structured uncertainty treatment, clear cost boundaries, and auditable assumptions.

IAEA guidance (e.g., TRS/TECDOC methodology) provides a standardized approach to economic evaluation and to the calculation of levelized electricity metrics such as the Levelized Discounted Electricity Generation Cost \cite{iaea-2001}. The IAEA methodology highlights the role of financial parameters (discount rate, inflation, escalation, and plant lifetime), and it emphasizes sensitivity analysis as a necessary complement to point estimates, particularly for long-lived infrastructure projects where economic conditions and technology maturity can evolve over time.

\subsection{Documents providing cost bases for key cost categories}
\label{subsec:cost_basis_docs}

To populate the cost accounts with defensible cost bases, we draw on fusion-specific cost documentation developed for the ARIES systems studies, alongside independent baselines for conventional power plants. The UCSD ``ARIES Cost Account Documentation'' report by Waganer provides historical and methodological context for fusion power plant costing and develops updated economic models suitable for forward-looking system studies \cite{Waganer2013}. It compiles and rationalizes costing information from earlier conceptual designs, documents escalation practices (e.g., use of macroeconomic deflators to translate historical estimates), and reviews the general cost-account structures originally defined in DOE-era guidance. It also addresses recurring issues in fusion costing such as spare parts, contingency, and safety assurance level assumptions.

Complementing this, Waganer’s ARIES project presentation on Cost Accounts 20 and 21 provides additional detail on the evolution of cost modeling for major plant structures and reactor/heat-island-adjacent equipment categories \cite{Waganer2007}. It motivates updates to legacy costing algorithms that were originally developed with limited documentation depth and highlights specific accounts that benefit from explicit definition and reporting (e.g., radioactive waste treatment, cryogenic systems, and plant maintenance equipment), which are also relevant to modern fusion facilities.

Finally, we use the National Energy Technology Laboratory (NETL) cost and performance baselines for fossil energy plants as an external reference for non-nuclear balance-of-plant components and for cross-technology benchmarking \cite{JamesCorrespondingAuthor2019}. The NETL baseline reports employ a systematic and transparent technical-economic approach across multiple plant configurations (e.g., IGCC, PC, and NGCC, with and without CO$_2$ capture). These baselines provide independently assessed cost and performance anchors for conventional plant subsystems and serve as a useful comparator set when evaluating fusion plant cost allocations and performance assumptions in a broader power-market context.

\section{Guiding philosophy}
Across all studies, we followed the ARIES-style workflow: begin with a physics-informed power balance, translate this into a feasible radial build subject to engineering constraints
(e.g., surface heat/power loading), then size the dominant driver systems (magnets, lasers, power supplies) and balance-of-plant (BOP). Assumptions and cost bases are stated explicitly in each
generated report, with BOP costs primarily informed by NETL baselines and augmented by additional sources as available to each concept and partner.

\section{Program Timeline and Study Objectives}

\subsection{Partners and stakeholders}
Key collaborators across the period included:
\begin{itemize}
  \item Bechtel (Desmond Chan, CTO)
  \item Decysive Systems (Ronald Miller)
  \item Lucid Catalyst (Eric Ingersoll)
  \item Princeton / PPPL (Mike Zarnstorff, lead POC) and PPPL domain experts (notably tritium handling and plant operations experience)
\end{itemize}

\subsection{2017: Capital-cost-focused pilot with Bechtel + Decysive Systems}
\subsubsection{Goals}
\begin{itemize}
  \item Establish an initial fusion costing workflow suitable for ARPA-E program needs.
  \item Produce capital cost estimates for representative fusion systems.
  \item Calibrate ARIES-derived scaling outputs against an experienced EPC (engineering, procurement, construction) perspective.
\end{itemize}

\subsubsection{Methodological basis}
\begin{itemize}
  \item Primary dependence on ARIES cost scaling relations (grounded in the UCSD-led body of work).
  \item Capital-cost scope only (no full indirect-cost model; limited treatment of O\&M and lifecycle effects).
\end{itemize}

\subsubsection{Role of Bechtel}
\begin{itemize}
  \item Provided calibration points and sanity checks using internal costing capability.
  \item Found rough agreement in BOP-related costs when compared to ARIES-derived scaling and internal benchmarks.
\end{itemize}

\subsubsection{Key outcomes}
\begin{itemize}
  \item Initial validation that ARIES-derived plant-level scalings were directionally consistent for several BOP categories.
  \item Identification of cost elements most sensitive to plant layout, buildings, and major equipment sizing.
\end{itemize}

\subsection{2019: BOP and indirect-cost deepening with Lucid Catalyst + Decysive Systems}
\subsubsection{Goals}
\begin{itemize}
  \item Separate and scrutinize BOP costs independent of EPC calibration.
  \item Expand emphasis on indirect costs and the costs outside the fusion power core.
  \item Identify actionable pathways to reduce dominant cost categories (buildings, turbine halls, heat exchangers, site works, construction management).
\end{itemize}

\subsubsection{Focus areas}
\begin{itemize}
  \item Indirect cost structure: engineering, construction management, owner's costs, contingency framing, schedule/cost-of-money implications.
  \item Design-for-cost approaches: standardized layouts, simplified buildings where feasible, and disciplined accounting structures.
  \item Modularization and centralized manufacturing: quantify learning opportunities and repeatability benefits.
\end{itemize}

\subsubsection{Key outcomes}
\begin{itemize}
  \item A comprehensive view of modularization feasibility and its potential impact on both direct and indirect costs.
  \item Clearer articulation that (for many concepts) non-power-core costs dominate total plant costs, motivating cost-out strategies beyond physics performance.
\end{itemize}

\subsection{2022--2023: Program-support role for ARPA-E BETHE and GAMOW (and revisits to ALPHA concepts)}
\subsubsection{Goals}
\begin{itemize}
  \item Apply learnings from 2017 and 2019 across a growing portfolio of ARPA-E fusion concepts.
  \item Provide rapid, concept-consistent costing analyses as a supporting-team function.
  \item Engage directly with private fusion companies, often delivering their first structured costing assessment.
\end{itemize}

\subsubsection{Activities}
\begin{itemize}
  \item Concept-by-concept cost driver identification for BETHE and GAMOW awardees.
  \item Revisit selected ALPHA concepts with updated assumptions and improved costing structure.
  \item Iterate power balance $\rightarrow$ radial build $\rightarrow$ driver system sizing $\rightarrow$ BOP translation for each architecture class:
  \begin{itemize}
    \item MFE: magnets and power supplies frequently dominant
    \item IFE: lasers and repetition-rate/target systems frequently dominant
    \item MIFE: hybrid driver dominance depending on configuration
  \end{itemize}
\end{itemize}

\subsubsection{Princeton/PPPL contribution (2022--2023)}
\begin{itemize}
  \item Mike Zarnstorff served as the lead POC at PPPL for the collaboration.
  \item Deepened treatment of tritium handling, leveraging decades of operational and safety experience from PPPL.
  \item Improved realism in fuel-cycle assumptions, containment approaches, and implications for plant systems and facilities.
\end{itemize}

\subsubsection{Key outcomes}
\begin{itemize}
  \item More accurate identification of cost drivers across diverse private-sector concepts.
  \item Better fidelity in tritium-related systems and associated facility/safety implications.
  \item More consistent cross-concept comparisons through harmonized assumptions and reporting.
\end{itemize}

\subsection{2023: Refactor to International Cost Accounts and Transition to Bottom-Up Power-Core Costing}
\subsubsection{Motivation}
By 2023, the program required a costing framework that was:
\begin{itemize}
  \item standardized (comparable across concepts and institutions),
  \item auditable (transparent cost basis and assumptions),
  \item extensible (new blankets, fuels, and driver technologies),
  \item and less dependent on legacy ARIES scalings for novel architectures.
\end{itemize}

\subsubsection{Cost-account standardization}
\begin{itemize}
  \item Refactored the cost code in accordance with the cost accounts originally proposed by the IAEA (2001),
  \item adopted by the GEN-IV Economics Modeling Working Group (EMWG) in 2007,
  \item and later adopted by EPRI (2024).
\end{itemize}

\subsubsection{Replacement of ARIES-derived scalings with bottom-up subsystem models}
Priority was given to bottom-up costing for major fusion-unique drivers and dominant plant cost contributors:
\begin{itemize}
  \item Magnets (including supporting structures and cryogenic implications as required)
  \item Lasers (including scaling with energy, repetition rate, efficiency, and facility implications)
  \item Power supplies (sized to electrical demand profiles and driver requirements)
  \item Power core components and key materials/volumes driven by radial build
\end{itemize}

\subsubsection{Physics-to-economics extensions}
\begin{itemize}
  \item Developed power balance and radial build capabilities for multiple fuel cycles:
  \begin{itemize}
    \item D--D, D--T, D--$^3$He, and p--$^{11}$B
  \end{itemize}
  \item Incorporated published blanket configurations and broadened blanket option coverage.
  \item Maintained explicit constraint handling (e.g., surface heat/power loading limits) to ensure economic outputs remained tied to feasible engineering envelopes.
\end{itemize}

\subsubsection{FECONs and pyFECONs: Tools, Releases, and Intended Use}
\subsection{FECONs (spreadsheet framework)}
\begin{itemize}
  \item Implemented the refactored accounts and bottom-up components in a spreadsheet tool named \textit{FECONs} (Fusion Economics code).
  \item Designed to demonstrate the level of fidelity required for credible fusion costing and to provide an auditable pathway from assumptions to totals.
\end{itemize}

\subsubsection{pyFECONs (open-source Python implementation)}
\begin{itemize}
  \item Converted FECONs to a Python script and released the open-source version as \textit{pyFECONs}.
  \item Published on GitHub as a framework illustrating structure, data flow, and required fidelity for fusion cost analysis.
  \item The open-source release intentionally emphasized transparency and extensibility over proprietary calibration.
\end{itemize}

\subsubsection{Clean Air Task Force supported extension (closed-source evolution)}
\begin{itemize}
  \item Subsequent development (post initial open-source release) was supported by the Clean Air Task Force.
  \item This branch is evolving into a closed-source version with a web interface and additional costing features, including expanded treatment of safety systems.
\end{itemize}

\subsection{2025 Application Example (Contextual, Post-ARPA-E Window)}
\subsubsection{LANL Plasma Jet Magneto-Inertial Fusion concept}
\begin{itemize}
  \item Applied the matured costing framework to a small number of systems, including the LANL Plasma Jet Magneto-Inertial Fusion concept.
  \item Released under an LA-UR number and served as a consolidation of the analysis fidelity achieved to date.
\end{itemize}

\subsection{Methodology (Common to All Studies)}
\subsubsection{Step 1: Physics-informed power balance}
\begin{itemize}
  \item Inputs range from analytic calculations to higher-fidelity modeling (including, where available, 3D MHD stability calculations).
  \item Outputs define required fusion power, driver power, recirculating power, and net electrical power targets.
\end{itemize}

\subsubsection{Step 2: Radial build and engineering constraints}
\begin{itemize}
  \item Translate physics requirements into geometry: plasma/target region, first wall, blanket, shield, structural allowances, and maintenance space.
  \item Enforce constraints (notably surface power loading and material/thermal limits), which strongly shape plant size and cost.
\end{itemize}

\subsubsection{Step 3: Driver system sizing (dominant cost drivers)}
\begin{itemize}
  \item MFE: magnets and associated systems (structure, cryogenics, power supplies).
  \item IFE: lasers, repetition-rate systems, and (where applicable) target-related handling/facilities.
  \item MIFE: hybrid dominance depending on architecture and operating point.
\end{itemize}

\subsubsection{Step 4: Balance-of-plant translation}
\begin{itemize}
  \item Map thermal power, conversion efficiency, and plant layout into turbine hall, heat exchangers, cooling/heat rejection, electrical plant, and buildings.
  \item BOP cost basis primarily referenced to NETL reports, supplemented by other sources as needed for the specific concept and data availability.
\end{itemize}

\subsubsection{Step 5: Indirect costs and total plant cost}
\begin{itemize}
  \item Evolve from early limited treatment (2017) to deeper, structured indirect cost consideration (2019 onward).
  \item Explicitly address the large influence of buildings, construction methodology, modularization, and learning on total plant cost.
\end{itemize}

\subsubsection{Step 6: Reporting, assumptions, and traceability}
\begin{itemize}
  \item Each report states assumptions, the cost basis for major accounts, and the conceptual maturity level.
  \item Outputs emphasize identification of dominant cost drivers and sensitivity to key design parameters.
\end{itemize}

\subsection{Evolution of the Costing Approach (Key Themes)}
\subsection{From scalings to bottoms-up}
\begin{itemize}
  \item 2017: ARIES-derived scalings used heavily for capital cost estimation with EPC calibration points.
  \item 2019: expanded indirect cost and modularization analysis; increased focus on non-power-core dominance.
  \item 2022--2023: portfolio-scale application with improved tritium realism and broader concept coverage.
  \item 2023: standards-aligned cost accounts; bottom-up costing for major drivers; multi-fuel and blanket option expansion.
\end{itemize}

\subsubsection{From single-study outputs to program infrastructure}
\begin{itemize}
  \item Transitioned from one-off analyses to a repeatable framework supporting multiple ARPA-E programs and private-sector concepts.
  \item Emphasized auditable structure (accounts), reproducibility (code), and extensibility (multiple fuels/blankets/driver types).
\end{itemize}

\subsection{Deliverables Summary}
\begin{itemize}
  \item 2017: Capital cost estimates with Bechtel calibration; ARIES-scaling-based foundation.
  \item 2019: BOP/indirect cost emphasis; modularization and centralized manufacturing implications.
  \item 2022--2023: BETHE/GAMOW support; ALPHA revisits; private-sector first-pass costing studies; tritium handling deepening with PPPL.
  \item 2023: Cost-account refactor (IAEA/EMWG/EPRI-aligned); bottom-up driver and power-core component models; multi-fuel and blanket coverage.
  \item FECONs: Spreadsheet-based framework capturing new fidelity requirements.
  \item pyFECONs: Open-source Python framework released on GitHub.
  \item CATF-supported branch: closed-source web-based evolution with added safety costing features.
\end{itemize}

\section{Level-1 (``10-level'') cost accounts: IAEA, GEN-IV (GIF/EMWG), and ARPA-E fusion costing}
In this work, ``10-level'' refers to the highest-level groupings in the cost code-of-accounts (COA)---the broad buckets that organize all lower-level accounts.
The GEN-IV (GIF/EMWG) COA expresses these as $\{10,20,30,40,50,60\}$; the IAEA TCIC COA uses closely corresponding groupings but (in the TRS-396 presentation)
enumerates most capital accounts as 21--41, 50--54, 60--62, and 70--72 ~\cite{gif-emwg-2007,iaea-2001}.

\subsubsection{IAEA NPP Total Capital Investment Cost (TCIC) account system (2001)}
The IAEA TCIC COA organizes capital investment into four main groupings (plus supplementary items), with the structure explicitly listed in Table~II of TRS-396.~\cite{iaea-2001}
\begin{itemize}
  \item \textbf{Base costs (Accounts 21--41).} The ``overnight'' plant construction cost before supplementary and financing.
  \begin{itemize}
    \item \textbf{Direct plant costs (Accounts 21--28):} buildings/structures at site (21); plant equipment in major systems (22--27); plus simulators (28).
    \item \textbf{Indirect services and site execution (Accounts 30--39):} supplier/A\&E engineering, design, project management (home office and site), construction supervision, construction labor, commissioning, trial test run, and construction facilities/materials.
    \item \textbf{Other base-cost services (Accounts 40--41):} staff training/technology transfer and related services (40), and housing/infrastructure (41).
  \end{itemize}
  \item \textbf{Supplementary costs (Accounts 50--54).} Transportation/insurance (50), spare parts (51), contingencies (52), insurance (53), and decommissioning costs if not treated elsewhere (54).
  \item \textbf{Financial costs (Accounts 60--62).} Escalation (60), interest during construction (61), and fees (62), applied to the capitalized investment scope.
  \item \textbf{Owner's costs (Accounts 70--72).} Owner's capital investment and services (70), escalation of owner's costs (71), and financing of owner's costs (72).
\end{itemize}

\subsubsection{GEN-IV (GIF/EMWG) COA (2007) for power generation plants}
The GIF/EMWG guidelines provide a harmonized COA and reporting template (Table~1.2) that is widely used for advanced nuclear economic studies and implemented in the G4ECONS model.~\cite{gif-emwg-2007}
At the ``10-level'' (highest level), the categories are:
\begin{itemize}
  \item \textbf{Account 10:} Capitalized pre-construction costs.
  \item \textbf{Account 20:} Capitalized direct costs.
  \item \textbf{Account 30:} Capitalized indirect services costs.
  \item \textbf{Account 40:} Capitalized owner’s costs.
  \item \textbf{Account 50:} Capitalized supplementary costs.
  \item \textbf{Account 60:} Capitalized financial costs.
\end{itemize}

\subsubsection{ARPA-E fusion costing (FECONs/pyFECONs) ``10-level'' accounts and how fusion maps into them}
For the ARPA-E fusion costing refactor (FECONs/pyFECONs), we adopted the GEN-IV (GIF/EMWG) 10-level capital COA as the top-level reporting structure
(for comparability with fission and other low-carbon technologies), while mapping fusion-unique subsystems into the appropriate direct-cost equipment/building accounts.~\cite{gif-emwg-2007}

\paragraph{Note on equivalence between IAEA and GIF ``10-level'' groupings.}
The GIF/EMWG 10-level accounts are a packaging of essentially the same scope elements represented in the IAEA TCIC COA, but expressed in a way that cleanly separates
pre-construction, direct, indirect services, owner’s, supplementary, and financial adders at the top level.~\cite{gif-emwg-2007,iaea-2001}

\paragraph{Included in the Appendix is a complete chart of accounts that has been used for the ARPA-E work.}  Note that the chart of accounts given is for a MFE system.  IFE and MIFE systems are almost identical, except for important ways such as the use lasers of pulsed power, and these major cost drivers are usually placed in 22.1.3.  However, there are important differences between each fusion concept outside of 22.1.3, and so the accounts and methodology are provided only as guidance: each power plant will be highly customized.

\section{Python fusion power plant costing code (pyFECONs example) overview}
\label{sec:pyfecons_overview}

\subsection{Implementation approach and cost-account realization}
\label{subsec:implementation_intro}

The \texttt{pyFECONs} example implements a sequential, physics-to-economics workflow in which each cost account is
realized from explicit intermediate plant quantities rather than imposed as a single top-down scaling.
The calculation begins by establishing a consistent \emph{power balance}, which determines gross and net electric
output and provides the capacity-normalization basis used throughout the costing (e.g., \$/kW$_e$ factors,
BOP sizing, and annual energy production for LCOE). With the net plant output fixed, the code proceeds to
a \emph{radial build} that defines the principal geometric dimensions of the fusion ``heat island'' (e.g., core,
blanket/shield regions, coil envelope, and bioshield). These geometry and layout results then drive the
sizing of the \emph{containment / heat-island building} and associated plant structures.

Next, the implementation sizes the major \emph{balance-of-plant (BOP)} systems---including turbines, generators,
condensers, heat rejection, and supporting electrical and auxiliary plant equipment---using engineering rules
and cost bases derived primarily from the NETL 2019 cost and performance baseline methodology. In practice,
this means that once thermal power and conversion efficiency are specified, the steam cycle and heat rejection
capacities are determined, turbine island equipment is scaled accordingly, and the remaining BOP subsystems are
scaled consistently with power-plant practice. Buildings and site facilities are then placed around the major
elements (heat island, turbine hall, electrical plant, cooling systems, service buildings), yielding an
internally consistent plant layout and building program that is commensurate with the sized equipment.

Only after these plant-level physical quantities are established does the code assemble the \emph{overnight
construction cost (OCC)} by summing the relevant direct cost accounts (pre-construction and construction
categories) and the explicitly sized structures and equipment. Finally, the code computes the remaining
overhead and capitalized indirect accounts (indirect services, owner’s costs, supplementary costs, and
financial costs) using predominantly bottom-up calculations and well-defined roll-ups, rather than embedding
such items implicitly within LCOE. This sequencing enforces traceability: each cost account is linked to
specific physical drivers (power, geometry, equipment size, and layout) and to documented cost bases, enabling
auditable updates as subsystem assumptions evolve.

\paragraph{Interdependencies and coupled constraints in power-core costing.}
A recurring theme in fusion power-plant costing is that the dominant power-core subsystems are strongly coupled, so that
a ``local'' design choice often propagates nonlinearly into several cost accounts. For example, magnet technology and
operating temperature influence not only the direct magnet cost (e.g., conductor, structure, and manufacturing) but also
the required power supplies, the cryogenic plant sizing and parasitic power, and the scope of safety and confinement
systems needed to manage cryogens and quench events; these knock-on effects can shift both CAPEX and the recirculating
power fraction that governs net output. Coolant selection is similarly interdependent with materials choices and safety:
for instance, FLiBe (and related molten salts) can be chemically aggressive and may require tighter materials constraints,
corrosion allowances, chemistry control systems, and inspection/maintenance provisions, while helium or water choices
shift heat-transfer equipment sizing, pumping power, and activation/contamination boundaries. Blanket thickness and
composition are also coupled to magnet size and shielding: higher-field or larger-radius magnet sets can relax neutron
shielding requirements but increase structural and cryogenic scope, whereas thinner blankets can reduce plant size while
potentially increasing required lithium enrichment (or other breeding enhancements) to maintain target tritium breeding
ratio (TBR), thereby coupling geometry to fuel-cycle and materials procurement assumptions. Beyond these subsystem
interdependencies, the framework enforces basic engineering constraints that shape feasible operating points and costs,
including surface power-loading limits on first-wall/structures, IFE chamber sizing to avoid first-wall melt limits, and
IFE chamber clearing (pumping) requirements that couple repetition rate and debris/ash removal to vacuum system power and
equipment size. Capturing these coupled dependencies is essential for credible TEA: it prevents double counting and
ensures that cost and performance sensitivities reflect physically and operationally consistent plant designs rather than
independent, unconstrained subsystem perturbations.

\subsection{High-level workflow}
At a high level, the script proceeds as follows:
\begin{enumerate}
  \item Define blanket/material options and global economic assumptions (construction time, NOAK/FOAK toggle, plant life, availability).
  \item Compute a plant power balance (fusion power, thermal power, gross and net electric power).
  \item Compute capital cost accounts:
  \begin{itemize}
    \item Account 10 (pre-construction),
    \item Account 20 (direct costs) including 21 (buildings) and 22--28 plant equipment subaccounts,
    \item Accounts 30/40/50/60 (indirect, owner, supplementary, financial).
  \end{itemize}
  \item Compute annualized costs (70 O\&M, 80 fuel, 90 annualized capital/finance).
  \item Compute LCOE and write results into LaTeX tables.
\end{enumerate}

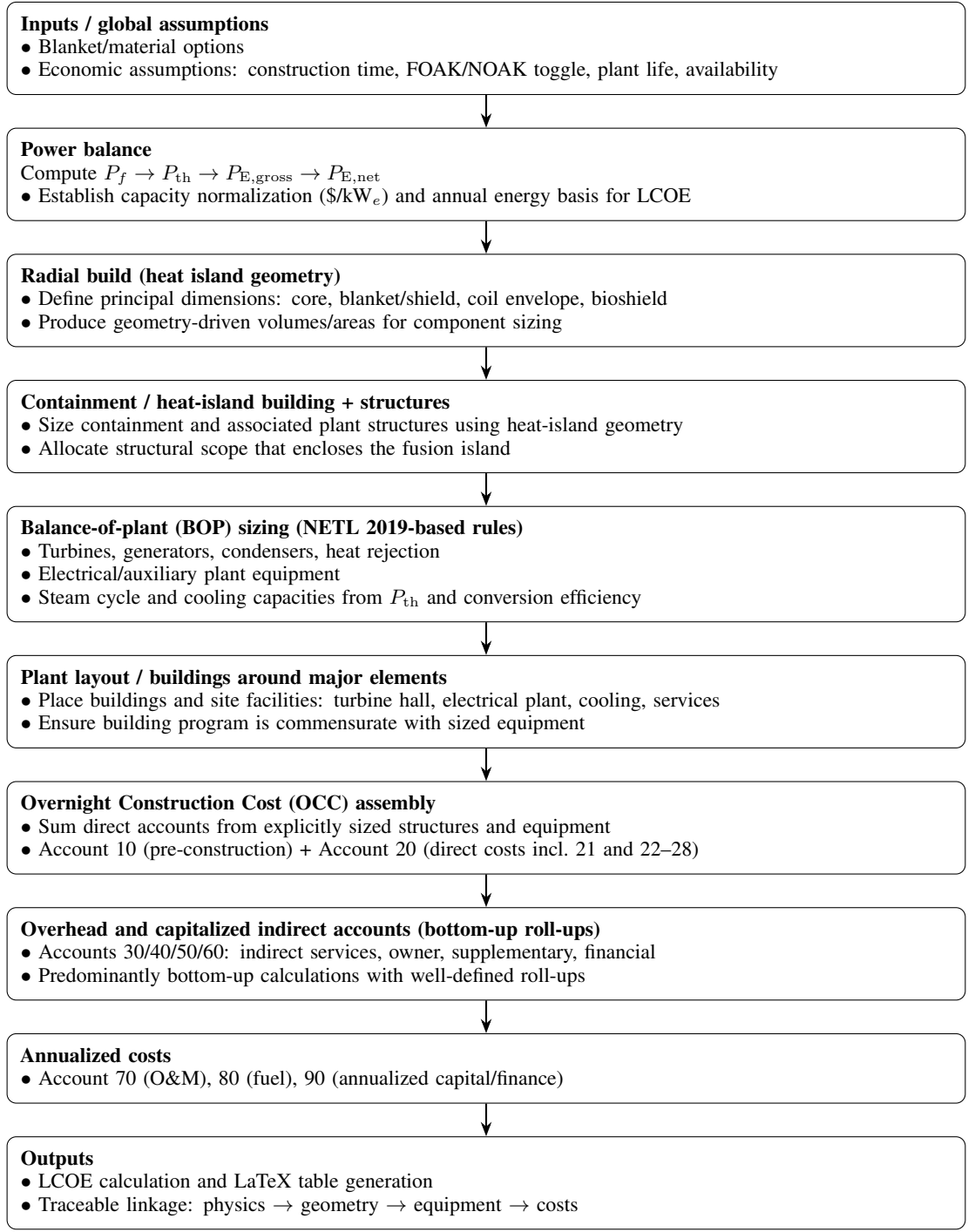
\begin{figure}[h!]
\centering
\begin{tikzpicture}[
  font=\small,
  node distance=5mm and 14mm,
  box/.style={rectangle, rounded corners, draw, align=left, inner sep=6pt, text width=0.86\linewidth},
  subbox/.style={rectangle, rounded corners, draw, align=left, inner sep=6pt, text width=0.82\linewidth},
  arrow/.style={-{Stealth[length=2.2mm]}, thick}
]

\node[box] (inputs) {\textbf{Inputs / global assumptions}\\
\(\bullet\) Blanket/material options\\
\(\bullet\) Economic assumptions: construction time, FOAK/NOAK toggle, plant life, availability};

\node[box, below=of inputs] (pb) {\textbf{Power balance}\\
Compute \(P_f \rightarrow P_{\mathrm{th}} \rightarrow P_{\mathrm{E,gross}} \rightarrow P_{\mathrm{E,net}}\)\\
\(\bullet\) Establish capacity normalization (\$/kW\(_e\)) and annual energy basis for LCOE};

\node[box, below=of pb] (radial) {\textbf{Radial build (heat island geometry)}\\
\(\bullet\) Define principal dimensions: core, blanket/shield, coil envelope, bioshield\\
\(\bullet\) Produce geometry-driven volumes/areas for component sizing};

\node[box, below=of radial] (contain) {\textbf{Containment / heat-island building + structures}\\
\(\bullet\) Size containment and associated plant structures using heat-island geometry\\
\(\bullet\) Allocate structural scope that encloses the fusion island};

\node[box, below=of contain] (bop) {\textbf{Balance-of-plant (BOP) sizing (NETL 2019-based rules)}\\
\(\bullet\) Turbines, generators, condensers, heat rejection\\
\(\bullet\) Electrical/auxiliary plant equipment\\
\(\bullet\) Steam cycle and cooling capacities from \(P_{\mathrm{th}}\) and conversion efficiency};

\node[box, below=of bop] (layout) {\textbf{Plant layout / buildings around major elements}\\
\(\bullet\) Place buildings and site facilities: turbine hall, electrical plant, cooling, services\\
\(\bullet\) Ensure building program is commensurate with sized equipment};

\node[box, below=of layout] (occ) {\textbf{Overnight Construction Cost (OCC) assembly}\\
\(\bullet\) Sum direct accounts from explicitly sized structures and equipment\\
\(\bullet\) Account 10 (pre-construction) + Account 20 (direct costs incl.\ 21 and 22--28)};

\node[box, below=of occ] (indirect) {\textbf{Overhead and capitalized indirect accounts (bottom-up roll-ups)}\\
\(\bullet\) Accounts 30/40/50/60: indirect services, owner, supplementary, financial\\
\(\bullet\) Predominantly bottom-up calculations with well-defined roll-ups};

\node[box, below=of indirect] (annual) {\textbf{Annualized costs}\\
\(\bullet\) Account 70 (O\&M), 80 (fuel), 90 (annualized capital/finance)};

\node[box, below=of annual] (lcoe) {\textbf{Outputs}\\
\(\bullet\) LCOE calculation and LaTeX table generation\\
\(\bullet\) Traceable linkage: physics \(\rightarrow\) geometry \(\rightarrow\) equipment \(\rightarrow\) costs};

\draw[arrow] (inputs) -- (pb);
\draw[arrow] (pb) -- (radial);
\draw[arrow] (radial) -- (contain);
\draw[arrow] (contain) -- (bop);
\draw[arrow] (bop) -- (layout);
\draw[arrow] (layout) -- (occ);
\draw[arrow] (occ) -- (indirect);
\draw[arrow] (indirect) -- (annual);
\draw[arrow] (annual) -- (lcoe);

\end{tikzpicture}
\caption{Sequential physics-to-economics workflow implemented in the \texttt{pyFECONs} example: power balance and net output establish the normalization basis; radial build and containment sizing determine heat-island geometry; NETL 2019-derived rules size the turbine island and BOP; plant buildings and layout are generated around major elements; direct costs are assembled into OCC; overhead and capitalized indirect accounts are rolled up bottom-up; annualized costs and LCOE are then computed and written to LaTeX tables.}
\label{fig:pyfecons_workflow}
\end{figure}

\subsection{Power balance model}
\label{sec:power_balance}

The script defines a steady-state power balance for a D--T-like energy partition, starting from a specified
fusion power $P_f$ (named \texttt{PNRL} in the code due to a legacy calculation of the fusion power using the NRL formulary \cite{NRLPlasmaFormulary2023} ). Alpha power $P_\alpha$ is computed from the D--T
energy split (3.52~MeV of 17.58~MeV), and neutron power is the remainder:  
\begin{align}
P_f &\equiv P_{\mathrm{NRL}}, \\
P_\alpha &= P_f \left(\frac{3.52}{17.58}\right), \\
P_n &= P_f - P_\alpha.
\end{align}

A neutron energy multiplier $M_N$ (named \texttt{MN}) is applied to neutron heating. Several plant internal
loads and fractions are defined, including:
\begin{align}
P_{\mathrm{aux}} &= P_{\mathrm{trit}} + P_{\mathrm{house}}, \\
P_{\mathrm{coil}} &= P_{\mathrm{TF}} + P_{\mathrm{PF}}, \\
P_{\mathrm{cool}} &= P_{\mathrm{TF,cool}} + P_{\mathrm{PF,cool}}, \\
P_{\mathrm{in}} &\equiv P_{\mathrm{INPUT1}} \quad \text{(auxiliary heating / driver input)}.
\end{align}

The code forms a thermal power expression (named \texttt{PTH}) that includes: neutron-multiplied neutron power,
alpha power, auxiliary heating input, and an additional term proportional to $(M_NP_n+P_\alpha)$ intended to
represent heat associated with primary pumping and subsystems (implemented as a fraction times the thermal conversion
efficiency):  
\begin{equation}
P_{\mathrm{th}} =
M_N P_n + P_\alpha + P_{\mathrm{in}}
+ \eta_{\mathrm{th}}\left(f_{\mathrm{PCPP}}\eta_p + f_{\mathrm{sub}}\right)\left(M_N P_n + P_\alpha\right),
\end{equation}
where $\eta_{\mathrm{th}}$ is the thermal-to-electric conversion efficiency, $f_{\mathrm{PCPP}}$ is the primary coolant
pumping fraction, $\eta_p$ is a pumping ``capture efficiency'' factor, and $f_{\mathrm{sub}}$ is a subsystem/control fraction.

Gross electric power is then:
\begin{align}
P_{\mathrm{th,e}} &= \eta_{\mathrm{th}} P_{\mathrm{th}}, \\
P_{\mathrm{DE}} &= \eta_{\mathrm{DE}} P_\alpha, \\
P_{\mathrm{E,gross}} &\equiv P_{\mathrm{ET}} = P_{\mathrm{th,e}} + P_{\mathrm{DE}},
\end{align}
with $\eta_{\mathrm{DE}}$ set to zero in the reviewed configuration.

The script defines a plant ``engineering gain'' $Q_{\mathrm{eng}}$ (named \texttt{QENG}) as a ratio of
``useful'' electric output to internal electric loads. With $\eta_{\mathrm{wp}}$ the wall-plug efficiency for auxiliary input power,
the implemented form is:  
\begin{equation}
Q_{\mathrm{eng}} =
\frac{\eta_{\mathrm{th}}\left(M_N P_n + P_{\mathrm{pump}} + P_{\mathrm{in}} + P_\alpha\right)}
{P_{\mathrm{coil}} + P_{\mathrm{pump}} + P_{\mathrm{sub}} + P_{\mathrm{aux}} + P_{\mathrm{cool}} + P_{\mathrm{cryo}} + P_{\mathrm{in}}/\eta_{\mathrm{wp}}},
\end{equation}
where the code sets
\begin{align}
P_{\mathrm{pump}} &= f_{\mathrm{PCPP}}\, P_{\mathrm{th,e}}, \\
P_{\mathrm{sub}} &= f_{\mathrm{sub}}\, P_{\mathrm{th,e}}.
\end{align}

The recirculating fraction is $f_{\mathrm{rec}} = 1/Q_{\mathrm{eng}}$ (named \texttt{REFRAC}), and net electric power is:  
\begin{equation}
P_{\mathrm{E,net}} \equiv P_{\mathrm{NET}}
= \left(1-\frac{1}{Q_{\mathrm{eng}}}\right) P_{\mathrm{E,gross}}.
\end{equation}

\paragraph{Extension to alternative fuels and direct energy conversion.}
The foregoing partition is ``D--T-like'' in that it assumes a dominant neutron channel with a fixed alpha fraction.
For alternative fuel cycles (e.g.\ D--D, D--$^3$He, and p--$^{11}$B), the power balance would require (i) replacing the
fixed $(3.52/17.58)$ alpha split with reaction-specific charged-particle and neutron energy fractions (including the
appropriate branching ratios for D--D), (ii) updating the neutron multiplier term to reflect the substantially reduced
(or, for some branches, redistributed) neutron heating, activation, and blanket multiplication pathways, and
(iii) revising auxiliary loads that are tightly coupled to fuel choice (e.g.\ tritium-breeding and processing loads may
decrease for D--$^3$He and p--$^{11}$B, while heating/driver requirements and wall-plug efficiency assumptions may shift due
to different optimal operating temperatures and confinement requirements). These changes also interact with the balance-of-plant:
a reduced neutron fraction can alter recoverable thermal power, blanket/coolant design choices, and heat-rejection requirements.
Finally, if a significant portion of charged-particle power can be converted directly to electricity (non-thermal
\emph{direct energy conversion}), the term \(P_{\mathrm{DE}}=\eta_{\mathrm{DE}}P_\alpha\) generalizes to the full charged-particle
channel, and the gross electric output becomes a hybrid of thermal conversion and direct conversion. In that case, the
recirculating-power accounting should avoid double-counting: parasitic electric loads should be subtracted from the combined
\(\eta_{\mathrm{th}}P_{\mathrm{th}} + P_{\mathrm{DE}}\) gross output, while thermal-side loads should remain in the thermal balance.

\paragraph{Account 21 (Buildings) scaling.}
Buildings and site structures are largely scaled linearly with gross electric power $P_{\mathrm{E,gross}}$
(\texttt{PET} in the code) using fixed coefficients (interpretable as \$/kW-type factors after escalation),
summed as:
\begin{equation}
C_{21} = \sum_{i} k_{21,i}\, P_{\mathrm{E,gross}} \;+\; C_{21,\mathrm{cont}},
\end{equation}
with an optional contingency term that is suppressed in NOAK mode.  

\paragraph{Account 22.1 (Fusion island / reactor equipment): bottom-up components.}
For several fusion-unique components, the script uses bottom-up volumetric/material models driven by a
radial build (thicknesses for vacuum vessel, first wall, blanket, shields, coil region, bioshield, etc.). Volumes
are computed from simplified torus/sphere geometry, and material costs are computed as:
\begin{equation}
C_{\mathrm{mat}} \approx
\frac{V\,\rho\,c_{\mathrm{raw}}\,m}{10^6},
\end{equation}
where $V$ is volume, $\rho$ is density, $c_{\mathrm{raw}}$ is a raw material unit cost, $m$ is a manufacturing
multiplier, and the factor $10^6$ converts to M\$ in the script. This structure is used, for example, for
first wall + blanket materials and for portions of shielding and special materials.  

\paragraph{Major cost driver: superconducting magnets (Account 22.1.3).}
A detailed magnet costing routine (\texttt{magnetsAll}) estimates REBCO tape length (or equivalent conductor usage),
computes superconductor material cost using a \$/kA$\cdot$m rate, adds copper/steel/insulation where applicable,
and applies manufacturing and structural multipliers. In simplified form, the implemented magnet cost logic is:
\begin{align}
C_{\mathrm{SC}} &\sim \frac{I_{\mathrm{tape}}}{10^3}\,\left(L_{\mathrm{tape}}\times 10^3\right)\,
\frac{c_{\mathrm{YBCO}}}{10^6}, \\
C_{\mathrm{Cu}} &\sim \frac{m_{\mathrm{Cu}}\,c_{\mathrm{Cu}}}{10^6}, \quad
C_{\mathrm{SS}} \sim \frac{m_{\mathrm{SS}}\,c_{\mathrm{SS}}}{10^6}, \\
C_{\mathrm{mag}} &=
m_{\mathrm{mfr}}\left(C_{\mathrm{SC}} + C_{\mathrm{Cu}} + C_{\mathrm{SS}} + C_{\mathrm{ins}}\right)
+ C_{\mathrm{struct}},
\end{align}
summed over coil sets (TF, CS, PF) and multiplied by coil counts. The code also includes an interpolation/extrapolation
path (via \texttt{LinearNDInterpolator} and linear regression) to auto-generate CICC parameters from a reference dataset.  

\paragraph{Treatment of IFE and MIF variants within Account 22.1.3 in the open-source code.}
While the MFE implementation treats superconducting magnets as the dominant 22.1.3 cost driver (as described above),
the GitHub versions of the framework generalize Account 22.1.3 to represent the \emph{primary driver technology} for the
architecture class. For inertial fusion energy (IFE) cases, 22.1.3 is populated by a detailed laser cost breakdown that
scales with delivered energy per pulse, repetition rate, wall-plug efficiency, optical train requirements, and associated
facility interfaces; the implementation separates laser subsystem elements (e.g., gain media / pump sources, optics, power
conditioning, thermal management, and integration allowances) to provide a transparent mapping from physics and driver
requirements to capital cost within the standardized account structure. For magneto-inertial fusion (MIF) concepts, the
same account is instead used to capture the pulsed-power driver train (and, where applicable, the associated magnet set),
including stored-energy scaling, pulse-forming networks, switches, transmission lines, and capacitor banks, with explicit
allowances for efficiency, repetition-rate constraints, and packaging into a plant-relevant driver module. In this way,
the open-source code preserves a consistent COA across concept classes while ensuring that the dominant, concept-defining
driver costs (lasers for IFE; pulsed power and often magnets for MIF; magnets for MFE) are represented at comparable
granularity and roll up cleanly into the 22.1 account totals.

\subsection{Cost-account structure implemented}
\label{sec:cost_accounts}

The code computes a capital cost total (named \texttt{C990000}) from top-level accounts:
\begin{equation}
C_{99} = C_{10} + C_{20} + C_{30} + C_{40} + C_{50} + C_{60},
\end{equation}
corresponding (respectively) to pre-construction, direct costs, indirect services, owner costs, supplementary costs,
and financial costs. Each account is computed as a sum of subaccounts with explicit variable names (e.g.,
\texttt{C210000} for buildings, \texttt{C220000} for reactor plant equipment, etc.).

\paragraph{Other notable direct-cost drivers.}
Additional drivers implemented with explicit scaling/estimation include:
\begin{itemize}
  \item power supplies (scaled from an ITER-based reference, with a learning credit factor),
  \item vacuum vessel and vacuum pumping (geometry-based volume, manufacturing factor, and pump-count scaling),
  \item cryogenic cooling power requirement estimation (Carnot-based COP fraction) and cost scaling from an ITER reference,
  \item supplementary heating cost scaling from ARIES/ITER reference \$/W values,
  \item divertor cost from volumetric tungsten,
  \item installation labor estimated from construction time and a scaling with machine size.
\end{itemize}
All of these contribute into $C_{22}$ and hence into $C_{20}$ and $C_{99}$.  

\subsection{Annualized costs and LCOE calculation}
\label{sec:lcoe}


The code forms three annualized cost terms:
\begin{itemize}
  \item \textbf{Annual O\&M cost (Account 70).}
  O\&M is computed using a lookup-based factor of 60~USD/(kW$_e$-yr), applied to net electric power
  $P_E \equiv P_{\mathrm{E,net}}$:
  \begin{align}
    C_{\mathrm{OM}}~[\mathrm{USD/yr}] &= 60 \; P_E \; (1000), \\
    C_{70}~[\mathrm{M\$\!/yr}] &= \frac{C_{\mathrm{OM}}}{10^{6}}
    = \frac{60 \; P_E \; (1000)}{10^{6}} .
  \end{align}

  \item \textbf{Annual fuel cost (Account 80).}
  The fuel cost is computed from fusion energy production to deuterium mass consumption using
  the D--T energy release (17.58~MeV per reaction) and the Joule-per-eV conversion factor.
  In the implemented form:
  \begin{align}
    m_D &= 3.342\times 10^{-27}\ \mathrm{kg}, \\
    u_D &= 2175\ \mathrm{\$/kg}, \\
    C_F~[\mathrm{USD/yr}] &=
    \frac{
      N_{\mathrm{mod}}\; P_f \; (10^{6})\; (3600)\; (8760)\; u_D\; m_D\; p_a
    }{
      17.58 \; (1.6021\times 10^{-13})
    }, \\
    C_{80}~[\mathrm{M\$\!/yr}] &= \frac{C_F}{10^{6}} .
  \end{align}
  Here $P_f$ is the fusion power (MW), $N_{\mathrm{mod}}$ is the number of modules,
  and $p_a$ is plant availability.

  \item \textbf{Annualized capital/financial cost (Account 90).}
  Annualized financial costs are computed as a fixed charge rate (capital return factor)
  multiplying the total capital cost:
  \begin{align}
    C_{\mathrm{AC}}~[\mathrm{USD/yr}] &= f_{\mathrm{cr}} \; (C_{99}\times 10^{6}), \\
    C_{90}~[\mathrm{M\$\!/yr}] &= \frac{C_{\mathrm{AC}}}{10^{6}} = f_{\mathrm{cr}} \; C_{99},
  \end{align}
  where $f_{\mathrm{cr}}$ is calculated following NETL 2019 conventions (constant-dollar FCR in the report).
\end{itemize}

\paragraph{LCOE as implemented in the report text.}
Contributors to LCOE are expressed as
\begin{equation}
  \mathrm{LCOE}\;[\$/\mathrm{MWh}] =
  \frac{
    C_{\mathrm{AC}} + \left(C_{\mathrm{OM}} + C_F\right)\,(1+y)^{Y}
  }{
    8760 \; P_E \; p_a
  },
\end{equation}
where $y$ is the annual fractional escalation factor applied to the recurring terms over a lifetime $Y$ (years),
$P_E$ is net electric power (MWe), and $p_a$ is plant availability.

\paragraph{Unit conversion.}
\begin{equation}
  \mathrm{LCOE}\;[\mathrm{cents/kWh}] = 0.1 \times \mathrm{LCOE}\;[\$/\mathrm{MWh}] .
\end{equation}

\subsection{Report generation and traceability}
The script includes helper functions that copy LaTeX templates (``Originals'') into a ``Modified'' directory and
overwrite placeholder strings with computed values. As a result, the cost breakdown, power balance tables,
and LCOE outputs are written into a consistent report structure, and the mapping from computed quantities
(e.g., $P_{\mathrm{E,net}}$, $C_{22.1.3}$ magnets) into cost-account totals ($C_{22}$, $C_{20}$, $C_{99}$) is explicit.  

\section{Extended models and optional analysis modules}
\label{sec:extended_models}

In the course of developing and applying the costing framework across multiple ARPA-E studies and subsequent customer
engagements, we have created a set of appendices and ``extra modules'' that extend the standard costing workflow without
requiring direct modification of the core code path. This architectural choice is intentional: maintaining a stable,
standards-aligned ``baseline'' implementation supports reproducibility and like-for-like comparisons across concepts,
while optional modules provide a controlled mechanism for exploring additional economic questions, deployment pathways,
and design-for-cost opportunities. In practice, these modules consume the same intermediate plant quantities and account
roll-ups produced by the standard workflow (power balance, plant sizing and layout, cost accounts, OCC/TCC, annualized
costs), then apply additional transformations, scenario logic, or valuation methods.

\subsection{Design-for-maintainability, reliability, and availability (RAM) cost-out modules}
A first family of extensions examines cost reductions achievable through explicit design-for-maintenance and reliability
engineering. These analyses explore how maintainability provisions, remote handling strategy, component segmentation,
spares philosophy, and planned replacement schedules can reduce downtime, improve availability, reduce corrective
maintenance burden, and shift lifecycle costs between capitalized scope (e.g., maintenance equipment, hot cells, shielding,
and access provisions) and recurring O\&M categories. The module framing supports trade studies in which increased upfront
CAPEX can be evaluated against reduced forced outage rates, reduced replacement labor, shorter maintenance durations, and
lower long-term operating costs.

\subsection{Retrofitting and repowering of existing thermal plants}
A second extension considers retrofitting (repowering) scenarios in which a fusion heat island replaces the combustion
heat source at an existing coal (or other thermal) station. The key methodological feature is that a large fraction of
the balance-of-plant (steam cycle, turbine island, electrical plant, cooling water/heat rejection infrastructure, and
portions of the buildings and site utilities) may be treated as inherited assets rather than newly purchased scope. This
provides a structured mechanism to evaluate FOAK deployment pathways that leverage existing infrastructure, potentially
obviating much of the BOP cost and shortening schedule risk, while explicitly accounting for retrofit-specific costs such
as integration engineering, demolition and site prep, new containment/heat-island building scope, and re-licensing or
permitting requirements.

\subsection{Materials scarcity, manufacturability, and supply-chain constraints}
A third family of extensions examines the economic implications of material choice under scarcity and manufacturability
constraints. Fusion concepts can be sensitive to the availability and processing routes of specialty materials, and some
candidate materials (e.g., beryllium) are not currently produced or traded in the quantities implied by large-scale
deployment. These modules are intended to (i) flag scarcity risk and potential price volatility, (ii) assess whether
fabrication and joining routes are compatible with the required component sizes and tolerances, and (iii) explore
substitutions or design adaptations that reduce reliance on constrained materials. The resulting outputs are expressed as
scenario-dependent adjustments to raw material unit costs, manufacturing multipliers, yield/scrap assumptions, and
availability of qualifying vendors, with transparent attribution to the affected cost accounts.

\subsection{Extended financial and economic valuation metrics (open-source add-ons)}
In addition to ``engineering-to-cost'' extensions, further modules have been developed in collaboration with customers and
clients and released as open source to support a broader set of financial valuation and decision metrics for fusion energy
projects. These add-ons allow users to compute, consistently and traceably from the standard costing outputs, a suite of
project finance and economic measures commonly used in utility planning, investment decision-making, and policy analysis:
\begin{itemize}[leftmargin=2em]
  \item Net Present Value (NPV), including explicit handling of discount rates and escalation assumptions;
  \item Total Life-Cycle Cost (TLCC);
  \item Revenue Requirements (e.g., required annual revenue stream to meet return and cost recovery targets);
  \item Levelized Cost of Energy (LCOE/LCOE variants) and related levelized-value concepts;
  \item Annualized Value (equivalent annual cost / equivalent annual value formulations);
  \item Internal Rate of Return (IRR) and Modified Internal Rate of Return (MIRR);
  \item Simple Payback Period and Discounted Payback Period;
  \item Benefit-to-Cost Ratios and Savings-to-Investment Ratios;
  \item Integrated Resource Planning / Demand-Side Management (IRP/DSM) ratio tests;
  \item Consumer and Producer Surplus metrics for welfare-oriented evaluations;
  \item Weighted Average Cost of Capital (WACC) calculation and its use within NPV and revenue-requirement analyses.
\end{itemize}

\paragraph{Rationale for modular extensions.}
These extended models are deliberately implemented as optional, composable modules rather than as edits to the baseline
code. This preserves the integrity of the standardized costing workflow (and the comparability it enables) while allowing
the framework to support diverse stakeholder questions: design-for-maintainability cost-out strategies, FOAK deployment via
retrofit pathways, material and manufacturability risk, and project finance metrics that go beyond point LCOE. In this way,
the methodology supports both a stable ``standard'' analysis mode and a growing ecosystem of targeted extensions that can
be applied when the decision context requires additional fidelity.


\section{Worked example: power balance, CAPEX, and LCOE}

This section reproduces a compact worked example of the costing workflow, including
(i) the steady-state power accounting, (ii) the resulting capital-cost roll-up by cost accounts,
and (iii) the levelized cost of electricity (LCOE) variable set and calculation.

\subsection{Power balance / power accounting}

Table~\ref{tab:power_accounting_worked} summarizes the steady-state power accounting used to establish
gross and net electric power, including auxiliary, compression/driver, and coolant-related recirculating loads.

\begin{table}[htbp]
\centering
\small
\caption{Worked example power accounting (all values in MW). This table fixes the thermal/electric
normalization basis for subsequent cost scaling (e.g., BOP sizing, \$/kW$_e$ factors, and annual net generation).}
\label{tab:power_accounting_worked}
\begin{tabularx}{\linewidth}{@{}l X r@{}}
\toprule
\textbf{Item} & \textbf{Description} & \textbf{Value [MW]} \\
\midrule
1.1 & Fusion power $P_f$ & 1290 \\
1.2 & Neutron multiplier $M_N$ & 1.05 \\
1.3 & Driver power $P_{\mathrm{in}}$ (compressor) & 30 \\
1.4 & House power $P_{\mathrm{house}}$ & 5 \\
1.5 & Wall-plug efficiency $\eta_{\mathrm{wp}}$ & 0.10 \\
1.6 & Coil power $P_{\mathrm{coil}}$ & 10 \\
1.7 & Auxiliary cooling power $P_{\mathrm{cool}}$ & 10 \\
1.8 & Cryogenic power $P_{\mathrm{cryo}}$ & 5 \\
1.9 & Thermal efficiency $\eta_{\mathrm{th}}$ & 0.35 \\
1.10 & Gross electric power $P_{\mathrm{E,gross}}$ & 828.12 \\
1.11 & $Q_{\mathrm{eng}}$ & 4.346 \\
3.1 & Pumping power $P_{\mathrm{pump}}$ & 289.84 \\
3.2 & Subsystems power $P_{\mathrm{sub}}$ & 31.99 \\
3.3 & Recirculating fraction $f_{\mathrm{rec}}$ & 0.230 \\
3.4 & Net electric power $P_{\mathrm{E,net}}$ & 636.75 \\
\bottomrule
\end{tabularx}
\end{table}

\subsection{CAPEX / cost-account roll-up}

Table~\ref{tab:capex_accounts_worked} reproduces the capital-cost account roll-up for this case, including
direct costs and the capitalized indirect/owner/supplementary/financial accounts. The total is the
Total Capital Cost (TCC), consistent with the costing structure used throughout the report.

\begin{table}[h!]
\centering
\small
\caption{Worked example CAPEX by cost accounts (updated costs and codes). The table reports the
account cost in MUSD and the percentage contribution to TCC. The rightmost columns provide the
MARS comparison values shown in the source report.}
\label{tab:capex_accounts_worked}
\begin{tabularx}{\linewidth}{@{}l X
S[table-format=4.0] S[table-format=3.0]
S[table-format=4.1] S[table-format=3.1]@{}}
\toprule
\textbf{Acct.} & \textbf{Cost account} &
{\textbf{Cost}} & {\textbf{\%}} &
{\textbf{MARS Cost}} & {\textbf{MARS \%}} \\
\midrule
20 & Direct Costs & 1538 & 71 & 7100.0 & 79.0 \\
21 & Structures/Site & 258 & 12 & 1200.0 & 13.0 \\
22 & Reactor Plant Equip. & 672 & 31 & 2700.0 & 29.0 \\
22.1 & Reactor Equip. & 441 & 20 & {} & {} \\
22.1.1 & First Wall \& Blanket & 150 & 7 & {} & {} \\
22.1.2 & Shield & 80 & 4 & {} & {} \\
22.1.3 & Magnets & 0 & 0 & 2200.0 & 24.0 \\
22.1.4 & Compression (Jets) & 25 & 1 & {} & {} \\
22.1.5 & Primary Structure & 0 & 0 & {} & {} \\
22.1.6 & Vacuum System & 20 & 1 & {} & {} \\
22.1.7 & Power Supplies & 111 & 5 & {} & {} \\
22.1.8 & Target Plasma & 10 & 0 & {} & {} \\
22.1.9 & Direct E. Conv. & 0 & 0 & {} & {} \\
22.1.11 & Assembly and installation & 5 & 0 & {} & {} \\
22.2 & Main Heat Transfer & 106 & 5 & {} & {} \\
22.3 & Auxiliary Cooling & 3 & 0 & {} & {} \\
22.4 & Rad. Waste Treat. & 5 & 0 & 140.0 & 1.5 \\
22.5 & Fuel Processing & 89 & 4 & 140.0 & 1.5 \\
22.6 & Other plant equipment & 8 & 0 & {} & {} \\
22.7 & Instrumentation and control & 21 & 1 & 160.0 & 1.8 \\
23 & Turbine Plant Equip. & 182 & 8 & 790.0 & 8.7 \\
24 & Electric Plant Equip. & 45 & 2 & {} & {} \\
25 & Misc. Plant Equip. & 32 & 1 & {} & {} \\
26 & Heat Rejection & 78 & 4 & {} & {} \\
27 & Special Materials & 3 & 0 & {} & {} \\
28 & Digital Twin & 5 & 0 & {} & {} \\
29 & Contingency & 0 & 0 & {} & {} \\
30 & Capitalized Indirect Service Costs (CISC) & 16 & 1 & 51.0 & 0.56 \\
40 & Capitalized Owner's Cost (COC) & 185 & 8 & 610.0 & 6.7 \\
50 & Capitalized Supplementary Costs (CSC) & 213 & 10 & 700.0 & 7.7 \\
60 & Capitalized Financial Costs (CFC) & 180 & 8 & 590.0 & 6.5 \\
\midrule
\multicolumn{2}{@{}l}{\textbf{Total Capital Cost (TCC)}} & \textbf{2173} & \textbf{100} & \textbf{9100.0} & \textbf{100.0} \\
\bottomrule
\end{tabularx}
\end{table}

\subsection{LCOE calculation}

The report’s LCOE definition is:
\begin{equation}
\mathrm{LCOE}\;[\$/\mathrm{MWh}] =
\frac{\mathrm{CAC} + (\mathrm{COM} + \mathrm{CF})(1+y)^{Y}}
{8760\,P_E\,p_a},
\label{eq:lcoe_worked}
\end{equation}
where $\mathrm{CAC}$ is the annual capital charge (TCC times fixed charge rate), $\mathrm{COM}$ is annual
operations and maintenance, $\mathrm{CF}$ is annual fuel cost, $y$ is the annual fractional increase applied
over plant life $Y$, $P_E$ is net electric power (MWe), and $p_a$ is plant availability.

Table~\ref{tab:lcoe_vars_worked} lists the variable values used for the worked case, and the reported
LCOE result.

\begin{table}[htbp]
\centering
\small
\caption{Variables used in the worked LCOE calculation. These inputs, together with
Eq.~\eqref{eq:lcoe_worked}, produce the reported LCOE of 55.1~\$/MWh (5.5~c/kWh).}
\label{tab:lcoe_vars_worked}
\begin{tabularx}{\linewidth}{@{}l X r@{}}
\toprule
\textbf{Quantity} & \textbf{Description} & \textbf{Value} \\
\midrule
$P_E$ & Net electric power (MW) & 636.748 \\
$p_a$ & Plant availability (-) & 0.9 \\
$y$ & Inflation / escalation (-) & 0.025 \\
$Y$ & Plant lifetime (years) & 30 \\
CAC & Capital cost (M\$/annum) & 195.6 \\
COM & O\&M cost (M\$/annum) & 38.2 \\
CF & Fuel cost (M\$/annum) & 1.0 \\
\midrule
\multicolumn{2}{@{}l}{\textbf{Reported LCOE}} & \textbf{55.1 \$/MWh (5.5 c/kWh)} \\
\bottomrule
\end{tabularx}
\end{table}

\section{Discussion: Relationship to Prior Power-Plant and Fusion Costing Methodologies}
\label{sec:discussion_prior_methodologies}

The ARPA-E fusion costing capability developed from 2017--2024 can be understood as a deliberate synthesis of (i) the
\emph{standards-aligned} nuclear power-plant code-of-accounts tradition (IAEA / GIF-EMWG / EPRI), (ii) the
\emph{physics-to-plant} systems-study workflow pioneered in fusion by ARIES and related studies, and (iii) an
\emph{EPC- and manufacturability-aware} cost-reduction framing (layout, constructability, modularization, learning)
motivated by both nuclear experience and ARPA-E program needs. In contrast to many earlier fusion costing studies that
either (a) reported only a capital-cost point estimate or (b) embedded important cost categories implicitly within an LCOE
formula, the ARPA-E workflow emphasized explicit account realization and traceability from physics assumptions to
plant-level quantities, subsystem sizing, and rolled-up accounts, ultimately enabling auditable updates and
portfolio-consistent comparisons.

\subsection{Contrast with GEN-IV / IAEA-style economic methodologies}
The GIF Economics Modeling Working Group (EMWG) guidelines were developed to support consistent economic evaluation across
Generation IV nuclear systems, including a harmonized chart of accounts (COA), reporting conventions, and explicit
attention to the distinction between top-down versus bottom-up cost estimating within an integrated economic modeling
workflow \cite{gif-emwg-2007,iaea-2001,epri-2024}. In the GEN-IV context, the COA primarily serves as a standardized
container for plant structures and systems, allowing results to be compared across designs, vendors, and national
programs.

The ARPA-E methodology adopts this same \emph{standards function} of the COA, but its main extension lies in how
fusion-specific subsystems are \emph{mapped into} and \emph{realized within} those accounts. Rather than treating fusion as
a small perturbation to a conventional nuclear plant, the ARPA-E implementation explicitly identifies the dominant
fusion-unique drivers (e.g., magnets, lasers, pulsed power, power supplies, and power-core components) and replaces legacy
single-factor scalings with bottom-up or semi-analytic subsystem models where those drivers dominate total cost.
A second practical distinction is emphasis on conventional balance-of-plant (BOP) engineering translation: conventional
subsystems (steam cycle, turbine island, heat rejection, electrical plant, and buildings) are sized consistently with
power-plant practice and anchored to transparent external baselines, ensuring that non-fusion plant elements remain
internally consistent as heat-island assumptions evolve.

Finally, GEN-IV/IAEA methodologies are typically deployed for concepts whose licensing basis, construction methods, and
project execution pathways are relatively well understood. The ARPA-E program work therefore placed particular emphasis on
indirect costs, construction methodology, and design-for-cost strategies, not merely as adders but as \emph{controllable
design outcomes}. This theme became central after early benchmarking exercises underscored that BOP, layout, and indirects
can dominate plant cost even when the fusion island is aggressively optimized.

\subsection{Contrast with ARIES and related fusion systems studies}
ARIES established a durable methodological template for fusion power-plant studies: couple a physics-informed power
balance to an engineering-constrained radial build, generate a conceptual plant configuration, and estimate costs using a
mix of subsystem models and scaling relations. The ARPA-E support work explicitly follows this ARIES-style workflow as the
guiding backbone across studies, while adapting it for broader portfolio use and for improved accounting traceability.

The ARPA-E methodology departs from classic ARIES practice in several important ways:
\begin{enumerate}
\item \textbf{Portfolio-consistent comparability across architectures.}
ARIES studies typically optimize and report within a given magnetic-fusion class, whereas ARPA-E had to support a rapidly
expanding portfolio spanning MFE, IFE, and MIF concepts with heterogeneous drivers, maturity levels, and plant integration
assumptions. The costing workflow therefore evolved to preserve a common COA while allowing concept-defining drivers to be
represented with comparable fidelity (e.g., magnets for MFE; lasers for IFE; pulsed power and sometimes magnets for MIF),
enabling more uniform cross-concept comparisons.

\item \textbf{EPC calibration and constructability emphasis.}
A distinguishing feature of the ARPA-E timeline is the explicit engagement of EPC-oriented validation and sanity checks,
which highlighted sensitivities to layout, buildings, and major equipment sizing and materially influenced later emphasis
on BOP, indirects, and design-for-cost.

\item \textbf{Explicit indirect-cost accounting and cost-reduction pathways.}
The ARPA-E workflow elevated indirect costs and execution pathways (construction supervision, project management, owner’s
costs, schedule effects, and cost-of-money implications) as first-class objects for sensitivity studies and cost-out
strategy. This emphasis reflects the recognition that non-power-core costs often dominate total plant cost for many fusion
concepts.

\item \textbf{Standards-aligned refactor and bottom-up replacement of legacy scalings.}
By 2023 the ARPA-E toolchain was refactored to align with the IAEA--GIF/EMWG--EPRI COA lineage, and key ARIES-derived
scalings were replaced with bottom-up subsystem models for dominant fusion cost drivers, coupled explicitly to power
balances and engineering-constrained radial builds. This shift moves the ARIES integrated workflow into a more auditable,
standards-aligned accounting envelope while increasing granularity where fusion cost uncertainty is highest.
\end{enumerate}

\subsection{Relation to LIFE and other inertial-fusion costing studies}
The LIFE program and related inertial-fusion energy studies provide an important comparison point because they developed
integrated systems models that include cost and performance scaling relationships for major subsystems, including driver
(laser), target systems, chamber, heat transfer, power conversion, and buildings \cite{life-ife}. The ARPA-E methodology is
consistent with LIFE in treating the driver as a dominant, concept-defining cost element that must be modeled explicitly
and traced to physics and operating requirements.

Where ARPA-E differs is principally structural and programmatic: (i) it enforces a standardized COA mapping across
MFE/IFE/MIF so that the dominant driver appears in a consistent account context (e.g., as the main contributor to 22.1.3
for IFE or MIF variants), (ii) it integrates an EPC- and externally anchored BOP sizing step to maintain conventional-plant
realism even as fusion island assumptions change, and (iii) it emphasizes explicit indirect-cost roll-ups and auditable
cost bases rather than concept-specific adders that are difficult to compare across architectures.

\subsection{Prior fusion costing analyses: Sheffield, Generomak, and the ESECOM-era tradition}
A useful historical antecedent that merits explicit acknowledgement is the set of generic fusion
physics/engineering/economic models developed in the 1980s, particularly the \emph{Generomak} model associated with
Sheffield and collaborators and later used in ESECOM-era analyses \cite{sheffield-generomak} (and updated considerably recently \cite{sheffield-milora-2016}). Generomak was explicitly
designed as a consistent framework for evaluating the economic viability of alternative magnetic fusion reactors using a
generic model of physics performance, engineering constraints, and plant economics. This tradition strongly parallels the
modern ``systems code'' ethos: economic conclusions are only meaningful when the physics, engineering, and plant
integration assumptions are internally consistent.

The ARPA-E methodology can be seen as a descendant of this systems-code lineage, but updated in three ways that reflect
contemporary needs. First, it adopts a modern, internationally recognizable COA structure (IAEA/GIF/EPRI-aligned), which
improves comparability with fission and with regulated power-plant accounting practices. Second, it reflects present-day
understanding that indirect costs and construction execution can dominate outcomes; hence modularization, centralized
manufacturing, and learning are treated as explicit levers rather than qualitative commentary. Third, it explicitly targets
auditable traceability from physics quantities to layout, BOP sizing, and account roll-ups, motivated by the needs of a
portfolio program supporting diverse private-sector concepts.

\subsection{Summary of distinguishing characteristics}
In summary, the ARPA-E methodology is not a rejection of GEN-IV/IAEA economic guidance nor of ARIES-style fusion studies;
rather, it is an operational synthesis tailored to portfolio decision-making and credibility requirements:
\begin{itemize}
  \item It \emph{inherits} the physics-to-plant integrated workflow of ARIES and earlier systems-code traditions (including
  Generomak/ESECOM), while increasing traceability and driver granularity.
  \item It \emph{adopts} the GEN-IV/IAEA/EPRI COA lineage as a standards-aligned reporting and comparability layer, while
  mapping fusion-unique subsystems into accounts in a way that preserves cross-architecture consistency.
  \item It \emph{extends} conventional practice by anchoring BOP and buildings to transparent external baselines and by
  treating indirect costs, execution pathways, and learning as central cost drivers that must be examined explicitly.
\end{itemize}
These choices reflect the central thesis that fusion economic credibility requires not only better physics and better
components, but also better accounting: cost drivers must be explicit, comparable, and auditable as subsystem assumptions
evolve and as concepts progress toward pilot-plant design.

\subsection{Forward-looking: the next phase of fusion power-plant costing and analysis}
\label{subsec:future_costing}

Fusion power-plant costing is rapidly transitioning from document-centric, point-design reporting toward
\emph{data-rich, optimization-driven, and provenance-complete} digital workflows in which economics is evaluated
concurrently with geometry and physics/engineering feasibility. A clear emerging pattern is the coupling of integrated
multi-physics plant models to multi-objective optimization loops, with increasing reliance on machine-learning surrogates
to make exploration of high-dimensional design spaces computationally tractable and to enable publication of Pareto
trade spaces rather than single ``best'' designs. \cite{Woodruff2026DesignOptimizationReview}

Recent work by nTtau Digital exemplifies this direction by embedding costing directly into an automated co-design loop:
parametric CAD geometry generation (plasma boundary, coils, structures, and plant integration) feeds an automated
multi-physics simulation stack (electromagnetics, structural mechanics, thermal analysis, and optional neutronics), and
\emph{as each component is instantiated}, its extracted quantities (e.g., mass, volume, surface area, power ratings, and
complexity measures) are mapped into a standardized code-of-accounts. This produces an auditable ``cost build file''
per design point that aggregates component estimates into subsystem costs and plant-level totals, with complete
provenance capture (inputs, code versions, solver settings, and outputs) and robust workflow orchestration across HPC
resources. \cite{nTtauDigital2026Workflow}

A second key development is the maturation of surrogate-assisted optimization as a practical mechanism for moving from
hundreds of evaluations to tens of thousands (or more), including ``surrogate-in-the-loop'' strategies in which high-fidelity
simulations are used selectively for verification and to resolve regions of high surrogate uncertainty. \cite{nTtauDigital2026Workflow}
This is particularly important for stellarator and other geometry-flexible concepts, where the dominant cost of evaluation
often comes from repeated equilibrium/coil computations and downstream constraint checking, and where constraints are
intrinsically geometric (coil curvature, clearances, maximum field, maintainability). \cite{nTtauDigital2026Workflow}
In parallel, broader private-sector practice is converging toward integrated whole-plant environments that evaluate
plasma, magnets, structures, blankets, power conversion, and \emph{cost} in a single workflow, then generate Pareto-optimal
sets using evolutionary multi-objective optimizers. \cite{Woodruff2026DesignOptimizationReview}

Looking forward, these developments suggest three concrete directions for fusion costing and analysis. First, \emph{cost
models will increasingly be treated as first-class citizens of the plant optimization loop}, rather than post-hoc adders:
COA-aligned costing will be triggered by the same evolving CAD and systems models that enforce feasibility constraints,
thereby making the cost model inherently geometry- and integration-aware (including building placement/layout effects and
BOP routing penalties). \cite{nTtauDigital2026Workflow} Second, the community is likely to shift from deterministic point
estimates to \emph{risk-informed and uncertainty-aware} economic results, where uncertainty propagation and probabilistic
constraint violation (or robustness margins) are incorporated directly into the optimization objectives and the resulting
Pareto fronts. \cite{Woodruff2026DesignOptimizationReview} Third, as global surrogates mature, the field will move toward
multi-fidelity ecosystems in which (i) standardized account-based costing and LCOE models remain the reporting backbone,
but (ii) the dominant technical drivers (transport, exhaust, coil engineering, maintenance, and manufacturing complexity)
are learned from increasingly large datasets generated by automated workflows, enabling faster design iteration and more
transparent comparison across architectures. \cite{nTtauDigital2026Workflow,Woodruff2026DesignOptimizationReview}

In this framing, the long-term role of standards-aligned codes of accounts is strengthened rather than diminished: the COA
provides the durable interface between evolving physics/engineering models and externally interpretable economic outputs.
The main shift is operational: fusion costing becomes a continuously evaluated, dataset-producing element of an integrated
digital design process, enabling both auditable ``design $\rightarrow$ cost build'' traceability and systematic exploration of
cost/performance/risk trade spaces at scales that were previously impractical. \cite{nTtauDigital2026Workflow}

\section{Further work}
\label{sec:further_work}

The costing framework described here is intended to be extensible: its value increases as additional
physics, engineering, safety, and supply-chain realism are incorporated in ways that preserve the same
standards-aligned chart-of-accounts mapping and auditability. We highlight several high-leverage directions
for further work.

\subsection{Integrating fusion safety into plant-level costing and design iteration}
A priority extension is a tighter integration of fusion safety analysis into the costing workflow, informed
by the ongoing work of the CATF IWG on fusion safety. The central methodological opportunity is to treat
safety not as a post-hoc checklist but as a \emph{design-coupled cost driver}: confinement strategy,
material selection, activation products, detritiation requirements, accident analysis assumptions, and
regulatory posture each impose tangible implications for facility layout, ventilation and detritiation
capacity, waste handling, remote maintenance provisions, and operational staffing. In practice, this suggests
a future workflow in which the safety case and its key bounding assumptions (e.g., credible release
fractions, confinement barriers, tritium inventory limits, waste classification approach, and emergency
planning envelope) are captured as explicit model inputs that (i) drive additions or modifications to
specific accounts (particularly buildings/structures, radioactive waste treatment, fuel handling, auxiliary
systems, instrumentation and control, and owner’s costs), and (ii) propagate into indirect costs through
construction complexity, QA/QC requirements, commissioning, and licensing scope. A structured integration of
the CATF IWG fusion safety thinking would therefore enable consistent sensitivity studies on the cost impact
of safety posture choices across concept classes, while improving the credibility and completeness of the
plant-level cost basis.

\subsection{Power-flow visualization and traceability via Sankey diagrams}
A second tractable improvement is richer visualization of the internal power accounting and its relationship
to cost drivers. The current workflow already computes the major terms in the plant power balance
(fusion power partition, thermal conversion, recirculating loads, and net electric output), but the results
are typically reported as tables. Sankey diagrams provide a natural complement by conveying the full
\emph{flow structure} of power: how fusion energy partitions into neutron and charged-particle channels; how
recoverable heat moves through the primary heat-transfer system; how gross electrical output is reduced by
auxiliary systems, coolant pumping, cryogenics, and driver wall-plug losses; and how these internal demands
determine net delivery to the grid. Such diagrams are especially useful when comparing fuel cycles (D--T,
D--D, D--$^3$He, p--$^{11}$B) and direct energy conversion options, where the qualitative flow structure
changes. Embedding Sankey generation as a first-class output of the code would improve interpretability for
non-specialists, strengthen internal consistency checks (e.g., preventing double counting between thermal and
direct conversion channels), and provide a more transparent bridge between recirculating power fractions and
the sizing/cost of the subsystems that create those loads.

\subsection{Materials cost bases and the challenge of NOAK assumptions}
A third major area for further work is the state of the art in materials costing for fusion-relevant
materials and purity specifications. The present framework primarily reports Nth-of-a-kind (NOAK) costing,
implicitly assuming that learning, supply-chain maturation, and industrialization have already reduced unit
costs toward steady-state values. This assumption is appropriate for long-term economic comparisons but it
creates a practical gap: for several fusion-relevant materials, \emph{transparent NOAK cost bases are not
readily available}, particularly at the purity levels required by some fuel-cycle, breeder, coolant, and
materials-compatibility constraints. A concrete example is lithium: while commodity lithium pricing is widely
reported, the marginal cost and market availability of \(\ge\)99.99\% purity lithium (and isotopically tailored
lithium, where applicable) remain poorly documented in open sources, and similar gaps exist for beryllium,
high-purity tungsten products, specialized steels, coatings, and certain ceramics. Addressing this gap will
likely require a hybrid approach that combines (i) vendor engagement and anonymized price curve development,
(ii) explicit specification of purity and form factor (ingots, granules, salts, alloys), (iii) treatment of
yields and scrap/recycle assumptions, and (iv) explicit scenario analysis separating FOAK procurement from NOAK
steady-state production. In the costing framework, such work would translate into improved raw-material unit
cost inputs, manufacturing multipliers that reflect realistic fabrication routes, and uncertainty bounds on
materials-dominated accounts (notably first wall/blanket and special materials) that can be propagated into
LCOE and total capital cost uncertainty.

\subsection{Summary}
Together, these extensions---(i) safety-informed account realization aligned with CATF fusion safety thinking,
(ii) Sankey-based power-flow visualization as a traceability tool, and (iii) improved materials cost bases
consistent with NOAK assumptions but grounded in realistic purity and supply-chain constraints---represent a
practical roadmap for increasing both the credibility and usefulness of fusion power-plant techno-economic
analysis as the field progresses toward pilot-plant design and eventual commercial deployment.

\section{Conclusion}
\label{sec:conclusion}

Fusion is entering a phase in which economic credibility and comparability are as important as physics performance: as concepts mature from laboratory demonstrations toward integrated pilot plants, stakeholders require cost estimates that are transparent, traceable to technical assumptions, and comparable across architectures. 
The work summarized in this paper responds to that need by consolidating ARPA-E-supported costing development from 2017--2024 into a standards-aligned, auditable framework that evolved from early capital-cost-focused, ARIES-style scaling studies into a physics-to-economics workflow that emphasizes explicit power balance, engineering-constrained geometry and layout, and bottom-up subsystem models for dominant fusion cost drivers. 

A central contribution is the explicit alignment of the account structure with the IAEA--GEN-IV EMWG--EPRI lineage, coupled to a computational implementation (FECONs and pyFECONs) that preserves traceability: each cost account is realized from intermediate physical plant quantities and documented cost bases, rather than being embedded implicitly in an LCOE result. 
This traceable mapping is what enables disciplined updating as subsystem assumptions evolve, and it supports like-for-like comparisons across fusion concepts and against other generation technologies. 

Looking forward, the same forces that are driving maturation of fusion design practice---increasing integration across physics, engineering, and plant-level systems modeling---are likely to drive a corresponding maturation in fusion costing. The next phase of credible fusion TEA should increasingly (i) couple costing to integrated design workflows (power balance $\rightarrow$ engineering constraints $\rightarrow$ layout $\rightarrow$ cost accounts), (ii) quantify uncertainty and risk in ways that are explicitly attributable to model inputs and constraint margins, and (iii) support rapid iteration across architectures and operating points without sacrificing auditability. The evolution of this capability into open (pyFECONs) and closed-source, interface-driven implementations reflects that trajectory, as the tooling shifts from one-off study artifacts toward reusable infrastructure that can support broader stakeholder engagement and higher-frequency design iteration.

\section*{Acknowledgements}

This work was performed with support from ARPA-E across the 2017--2024 period, during which the costing methodology evolved from early capital-cost-focused studies to a standards-aligned, auditable capability supporting portfolio-scale application. 
Subsequent development (post open-source release) was supported by the Clean Air Task Force (CATF), including continued evolution of the toolchain toward a closed-source branch with a web interface and expanded costing features. 

I gratefully acknowledge the collaborators and stakeholders who contributed materially to the development and validation of the approach across its stages, including Bechtel, Decysive Systems, Lucid Catalyst, and Princeton/PPPL (including contributions informed by PPPL operational and tritium-handling expertise). I also thank my peers and contributors across the ARPA-E support effort and subsequent extensions.

For implementation contributions, I acknowledge Alex Higginbottom for the \texttt{pyFECONs} implementation in Python/Colab, and Chris Raastad for implementing the Python library and the web-interface used in the CATF (closed-source) version of the code.

Finally, I acknowledge the CATF IWG on Fusion Cost Analysis, including Laila El-Guebaly, Alicia Durham, and Geoffrey Rothwell, for guidance and constructive feedback on costing structure, reporting conventions, and economic interpretation.

\newpage
\appendix
\section{Chart of Accounts}\label{app:chart_of_accounts}

\begin{longtable}{p{0.18\textwidth}p{0.78\textwidth}}
\caption{Chart of accounts (filtered to account labels / subsystem names, excluding cost-basis numeric rows).}\\
\hline
\textbf{Account} & \textbf{Description}\\
\hline
\endfirsthead
\hline
\textbf{Account} & \textbf{Description}\\
\hline
\endhead

10 & Pre-construction costs\\
11 & Accessory Electric Plant (bare erected)\\
12 & Site Permits\\
13 & Plant Licensing\\
14 & Plant Permits\\
15 & Plant Studies\\
16 & Plant Reports\\
17 & Other Pre-Construction Costs\\
19 & Contingency on Pre-Construction Costs\\

20 & Capitalized Direct Costs (CDC)\\
21 & Structures and Improvements\\
21.1 & \hspace*{0.8em}Site Preparation / Yard Work\\
21.2 & \hspace*{0.8em}Heat Island Building\\
21.3 & \hspace*{0.8em}Turbine Generator Building\\

22 & Heat Island Plant Equipment\\
22.1 & \hspace*{0.8em}Heat Island Components\\
22.1.1 & \hspace*{1.6em}First Wall and Blanket\\
22.1.2 & \hspace*{1.6em}Shield\\
22.1.3 & \hspace*{1.6em}Coils or Lasers or Pulsed Power\\
22.1.4 & \hspace*{1.6em}Supplementary Heating Systems\\
22.1.5 & \hspace*{1.6em}Primary Structure and Support\\
22.1.6 & \hspace*{1.6em}Vacuum System\\
22.1.7 & \hspace*{1.6em}Power Supplies\\
22.1.8 & \hspace*{1.6em}Electrodes or Plasma Guns\\
22.1.9 & \hspace*{1.6em}Direct Energy Convertor\\

22.01.11 & \hspace*{0.8em}Assembly and Installation Costs\\
22.01.11.01 & \hspace*{1.6em}First wall and blanket\\
22.01.11.02 & \hspace*{1.6em}Shield\\
22.01.11.03 & \hspace*{1.6em}Magnets\\
22.01.11.04 & \hspace*{1.6em}Auxiliary heating\\
22.01.11.05 & \hspace*{1.6em}Primary structure\\
22.01.11.06 & \hspace*{1.6em}Vacuum system\\
22.01.11.07 & \hspace*{1.6em}Power supplies\\
22.01.11.08 & \hspace*{1.6em}Divertor\\
22.01.11.09 & \hspace*{1.6em}Direct energy convertor\\

22.2 & \hspace*{0.8em}Main and Secondary Coolant\\
22.02.00.00 & \hspace*{1.6em}Main heat transfer and transport systems\\
22.02.01.00 & \hspace*{1.6em}Primary coolant system\\
22.02.01.01 & \hspace*{1.6em}Pumps and motor drives (modular \& nonmodular)\\
22.02.01.02 & \hspace*{1.6em}Piping\\
22.02.01.03 & \hspace*{1.6em}Heat exchangers\\
22.02.01.04 & \hspace*{1.6em}Tanks (dump, make-up, clean-up, trit., hot storage)\\
22.02.01.05 & \hspace*{1.6em}Clean-up system\\
22.02.01.06 & \hspace*{1.6em}Thermal insulation, piping \& equipment\\
22.02.01.07 & \hspace*{1.6em}Tritium extraction\\
22.02.01.08 & \hspace*{1.6em}Pressurizer\\
22.02.01.09 & \hspace*{1.6em}Other\\
22.02.03.00 & \hspace*{1.6em}Secondary coolant system\\
22.02.03.01 & \hspace*{1.6em}Pumps and motor drives (modular \& non-modular)\\
22.02.03.02 & \hspace*{1.6em}Piping\\
22.02.03.03 & \hspace*{1.6em}Heat exchangers\\
22.02.03.04 & \hspace*{1.6em}Tanks (dump, make-up, clean-up, trit., hot storage)\\
22.02.03.05 & \hspace*{1.6em}Clean-up system\\
22.02.03.06 & \hspace*{1.6em}Thermal insulation, piping \& equipment\\
22.02.03.07 & \hspace*{1.6em}Tritium extraction\\
22.02.03.08 & \hspace*{1.6em}Pressurizer\\
22.02.03.09 & \hspace*{1.6em}Other\\
22.02.04.00 & \hspace*{1.6em}Thermal storage system\\

22.03 & \hspace*{0.8em}Auxiliary Cooling Systems\\
22.03.04 & \hspace*{1.6em}Power Supply and Cooling System\\
22.03.04.01 & \hspace*{2.4em}Refrigeration\\
22.03.04.02 & \hspace*{2.4em}Piping\\
22.03.04.03 & \hspace*{2.4em}Fluid circulation driving system\\
22.03.04.04 & \hspace*{2.4em}Tanks\\
22.03.04.05 & \hspace*{2.4em}Purification\\
22.03.05 & \hspace*{1.6em}Other cooling systems\\

22.04 & \hspace*{0.8em}Radioactive Waste Treatment\\
22.04.01 & \hspace*{1.6em}Liquid waste processing and equipment\\
22.04.02 & \hspace*{1.6em}Gaseous wastes and off-gas processing system\\
22.04.03 & \hspace*{1.6em}Solid waste processing equipment\\

22.5 & \hspace*{0.8em}Fuel Handling and Storage\\
22.6 & \hspace*{0.8em}Other Reactor Plant Equipment\\
22.7 & \hspace*{0.8em}Instrumentation and Control\\

23 & Turbine Plant Equipment\\
24 & Electric Plant Equipment\\
25 & Miscellaneous Plant Equipment\\
26 & Heat Rejection\\
27 & Special Materials\\
28 & Digital Twin / Simulator\\
29 & Contingency on Direct Capital Costs\\

30 & Capitalized Indirect Service Costs (CISC)\\
31 & Field Indirect Costs\\
32 & Construction Supervision\\
33 & Commissioning and Start-up Costs\\
34 & Demonstration Test Run\\
35 & Design Services (Offsite)\\
36 & PM/CM Services (Offsite)\\
37 & Design Services (Onsite)\\
38 & PM/CM Services (Onsite)\\
39 & Contingency on Support Services\\

40 & Capitalized Owner’s Cost (COC)\\
41 & Staff Recruitment and Training\\
42 & Staff Housing\\
43 & Staff Salary-Related Costs\\
44 & Other Owner’s Costs\\
49 & Contingency on Owner’s Costs\\

50 & Capitalized Supplementary Costs (CSC)\\
51 & Shipping and Transportation Costs\\
52 & Spare Parts\\
53 & Taxes\\
54 & Insurance\\
55 & Initial Fuel Load\\
58 & Decommissioning Costs\\
59 & Contingency on Supplementary Costs\\

60 & Capitalized Financial Costs (CFC)\\
61 & Escalation\\
62 & Fees\\
63 & Interest During Construction (IDC)\\
69 & Contingency on Capitalized Financial Costs\\

70 & Annualized O\&M Cost (AOC)\\
71 & O\&M Staff\\
72 & Management Staff\\
73 & Salary-Related Costs\\
74 & Operations Chemicals and Lubricants\\
75 & Spare Parts\\
76 & Utilities, Supplies, and Consumables\\
77 & Capital Plant Upgrades\\
78 & Taxes and Insurance\\
79 & Contingency on Annualized O\&M Costs\\

80 & Annualized Fuel Cost (AFC)\\
81 & Refueling Operations\\
84 & Fuel\\
86 & Processing Charges\\
87 & Special Nuclear Materials\\
89 & Contingency on Annualized Fuel Costs\\

90 & Annualized Financial Costs\\
92 & Home Office Engineering Services\\

\hline
\end{longtable}


\end{document}